\documentclass[sigconf]{acmart}
\usepackage{dirtytalk}
\usepackage{enumitem}
\usepackage{multirow}
\usepackage{listings}
\usepackage{xcolor}
\usepackage{makecell}

\AtBeginDocument{
  }
\setcopyright{acmlicensed}
\copyrightyear{2025}
\acmYear{2025}
\acmDOI{XXXXXXX.XXXXXXX}

\acmConference[Koli Calling '25]{25th International Conference on Computing Education Research}{November 11--16, 2025}{Koli, Finland}
\acmISBN{978-1-4503-XXXX-X/2018/06}

\begin{document}

\title{GenAI Voice Mode in Programming Education}

\author{Sven Jacobs}
\orcid{0009-0000-5079-7941}
\affiliation{
  \institution{University of Siegen}
  \city{Siegen}
  \country{Germany}
}
\email{sven.jacobs@uni-siegen.de}

\author{Natalie Kiesler}
\orcid{0000-0002-6843-2729}
\affiliation{
   \institution{Nuremberg Tech}
   \city{Nuremberg}
   \country{Germany}
}
\email{natalie.kiesler@th-nuernberg.de}

\renewcommand{\shortauthors}{Jacobs, Kiesler}

\begin{abstract}
Real-time voice interfaces using multimodal Generative AI (GenAI) can potentially address the accessibility needs of novice programmers with disabilities (e.g., related to vision). Yet, little is known about how novices interact with GenAI tools and their feedback quality in the form of audio output. This paper analyzes audio dialogues from nine 9th-grade students using a voice-enabled tutor (powered by OpenAI's Realtime API) in an authentic classroom setting while learning Python. We examined the students' voice prompts and AI's responses (1210 messages) by using qualitative coding. We also gathered students' perceptions via the Partner Modeling Questionnaire. The GenAI Voice Tutor primarily offered feedback on mistakes and next steps, but its correctness was limited (71.4\% correct out of 416 feedback outputs). Quality issues were observed, particularly when the AI attempted to utter programming code elements. Students used the GenAI voice tutor primarily for debugging. They perceived it as competent, only somewhat human-like, and flexible. The present study is the first to explore the interaction dynamics of real-time voice GenAI tutors and novice programmers, informing future educational tool design and potentially addressing accessibility needs of diverse learners. 
\end{abstract}

\ccsdesc[500]{Social and professional topics~Computing education}
\ccsdesc[500]{Computing methodologies~Artificial intelligence}

\keywords{GenAI, Feedback, Multimodal, Advanced Voice Mode, Realtime}


\maketitle

\section{Introduction}

Generative AI (GenAI) and related tools have taken the world by storm~\cite{prather2023therobots:wgfull}, and computing educators are starting to integrate them into their classes, assessments, and curricula~\cite{prather_beyond_2024}. There is no doubt about the potential of GenAI tools, e.g., for generating code explanations~\cite{macneil2022experiences}, exercises~\cite{jacobs2025unlimited}, code fixes~\cite{phung2023generating}, or different types of feedback~\cite{jacobs_evaluating_2024, kiesler2023exploring, lohr2025you}, whereas feedback is considered a crucial factor for learning~\cite{hattietimperley2007}.
At the same time, however, programming novices may struggle using GenAI tools, as they may develop an illusion of competence, depend on the tools, and lack critical thinking~\cite{jost2024theimpact,prather_widening_2024,scholl_analyzing_2024}. More important are the concerns regarding the tools' accessibility issues~\cite{alshaigy2024forgotten}, which add to the challenges of novice learners of programming~\cite{becker20219what,luxtonreilly2018introductory}.

Voice Assist may constitute an interesting alternative for diverse learners (e.g., neurodiverse, with impaired vision, or various reading and writing skills) interacting with GenAI tools. For example, \citet{poddar_experiences_2024} integrated an interactive voice-response into a learning environment for visually impaired students, resulting in strong student engagement and collaboration. The adoption of this modality for GenAI feedback thus seems worth implementing and exploring. This is particularly true for multimodal LLMs, such as GPT-4o~\cite{openai_gpt-4o_2024}, which interprets and outputs voice, allowing for lower latency and interpretation of accents, emotions, and emphasis. Access to voice capabilities of GPT-4o~\cite{openai_gpt-4o_2024} is available since September 2024 in ChatGPT (\emph{Advanced Voice Mode}) and as API (\emph{OpenAI Realtime API}) since October 2024.
However, the real-time voice capabilities of these LLMs have not yet been subject to research in educational contexts. So, we aim to explore them with able-bodied individuals before introducing them to learners requiring additional assistance.

The \textbf{goal} of this case study is to explore \cite{yin_case_2017} the voice mode of a recent GenAI model, and how it is used and perceived by learners of programming in secondary school (9th grade). This work is guided by the following research questions: 
\begin{enumerate}
\item[RQ1] \textit{How do novice programmers interact via voiced prompts with a real-time GenAI voice tutor during programming tasks?}
\item[RQ2] \textit{How can the GenAI tutor's voiced responses be characterized?}
\item[RQ3] \textit{How do novice programmers overall perceive the GenAI Voice Tutor as a dialogue partner?}
\end{enumerate}
\noindent
The \textbf{contributions} comprise an exploration of students' interaction patterns with voice generation as part of the Tutor Kai, a characterization of the GenAI voice responses in terms of its answer type, feedback category, and quality (i.e., correctness and issues). Finally, we offer insights into the student perspective of the voice generation. Thereby, this work has implications for tool designers, educators and learners interested in using GenAI's voice mode.

\section{Related Work} \label{sec: Related Work}

Although GenAI and related tools can help students of programming, e.g., via individual feedback~\cite{lohr2025you, scholl2025students}, they are inherently limited, intransparent, and exclusive by design~\cite{alshaigy2024forgotten,kiesler2025therole}. GenAI models are designed by able-bodied individuals and represent their respective knowledge and language as part of the selected training data~\cite{bender2021onthedangers}, ~\cite{alshaigy2024forgotten}. 
\citet{alshaigy2024forgotten} identified deficits of GenAI tools related to their inclusivity, e.g., blind persons using ChatGPT experience difficulties in the interaction with detailed charts and graphs, color-coded information, and dynamic content~\cite{alshaigy2024forgotten}. General issues are non-inclusive language, and biased/racist outputs~\cite{apiola2024first,alshaigy2024forgotten}.

Educational research has shown the potential of Voice Assistants (VAs), indicating VAs can support learning in primary~\cite{butler_ok_2024} and higher education~\cite{essel_exploring_2025}. 
For example, \citet{sayago_voice_2021} explored Apple Siri and Google Assistant as learning companions for undergraduate CS students in two courses. They found these VAs incompetent, incapable of answering complex questions, and observed communication misunderstandings \cite{sayago_voice_2021}.
\citet{jaber_cooking_2024} found that VAs (Google Assistant) struggle with supporting tasks like cooking due to a lack of context awareness, leading to irrelevant responses. Simulating context awareness revealed that users interact more fluently~\cite{jaber_cooking_2024}. 
Integrating Large Language Models (LLMs) offers advancements in addressing these observed context and interaction challenges. \citet{mahmood_user_2025} showed that an LLM-powered VA (using an Alexa-ChatGPT pipeline) enhances conversational resilience by using contextual understanding to overcome many speech recognition errors.
Despite these improvements, LLM-based VAs can introduce new issues like verbosity and repetition~\cite{mahmood_user_2025}. 
Prior work~\cite{butler_ok_2024, essel_exploring_2025, jaber_cooking_2024, mahmood_user_2025} relied on distinct Speech-to-Text (STT), text processing, and Text-to-Speech (TTS) stages, causing latency and the inability to interpret accents and emphasis. Recent developments in multimodal LLMs like OpenAI's GPT-4o \cite{openai_gpt-4o_2024} (available as \emph{Advanced Voice Mode} in ChatGPT), however, hold promise to overcome these issues. Yet, multimodal GenAI's real-time voice features have not been researched in educational contexts.

Recent research of text-based student-GenAI dialogs with chatbots (e.g., ChatGPT, GitHub Copilot, and custom tools)~\cite{prather_beyond_2024} shows that students use GenAI for many purposes, including problem understanding, conceptual understanding, code generation, debugging, and documentation~\cite{scholl_how_2024}. \citet{scholl_analyzing_2024} identified diverse chat patterns, from students seeking quick solutions with minimal prompting (cf.~\cite{amoozadeh_student-ai_2024}) to those engaging in more extended dialogues using the tool as a \say{learning partner}. This aligns with research on GitHub Copilot users, who operate in either an \emph{acceleration mode} (knowing what to do and using GenAI to do it faster) or an \emph{exploration mode} (being unsure and using GenAI to investigate options)~\cite{barke_grounded_2023}.
\citet{alfageeh_prompts_2025} introduced a logic-based lens to analyze student prompts, finding that iterative refinement often correlated with shorter solution paths. At the same time, significant modifications mid-session could indicate struggle.
\citet{prather_widening_2024} found that while some novice students can use GenAI to accelerate their work, struggling students may experience existing metacognitive difficulties \cite{prather_metacognitive_2018} compounded by GenAI, potentially leading to an \say{illusion of competence}.
GenAI can also introduce GenAI-specific metacognitive challenges, e.g., being misled by the tool or interrupted by frequent, sometimes unhelpful, suggestions \cite{prather_widening_2024}. 
Efforts to mitigate these issues include developing tools with pedagogical guardrails and refining system prompts \cite{jacobs_evaluating_2024, liffiton_codehelp_2023, kazemitabaar_codeaid_2024,scholl2025students}.

While research on text-based GenAI tutors in programming education is expanding, real-time voice interactions of novice programmers with multimodal models remain a critical unexplored area, particularly as \citet{stone_exploring_2024} underscores the urgent need for more human-centered research on how \say{second-level students} engage with GenAI.
Our study addresses this gap by examining verbal student-GenAI interactions and perceptions of ninth-grade learners in secondary education at a public school in Germany. We also evaluate the capability of the underlying multimodal GenAI Model to generate voice feedback for novice programmers in an authentic learning context. Given the lack of accessibility, equity, and inclusion of recent GenAI tools~\cite{alshaigy2024forgotten} (in addition to their other challenges~\cite{prather_widening_2024}), exploring voice interactions may also benefit diverse learners and educators utilizing GenAI.

\begin{figure}[tb]
    \centering
    \includegraphics[width=1\linewidth]{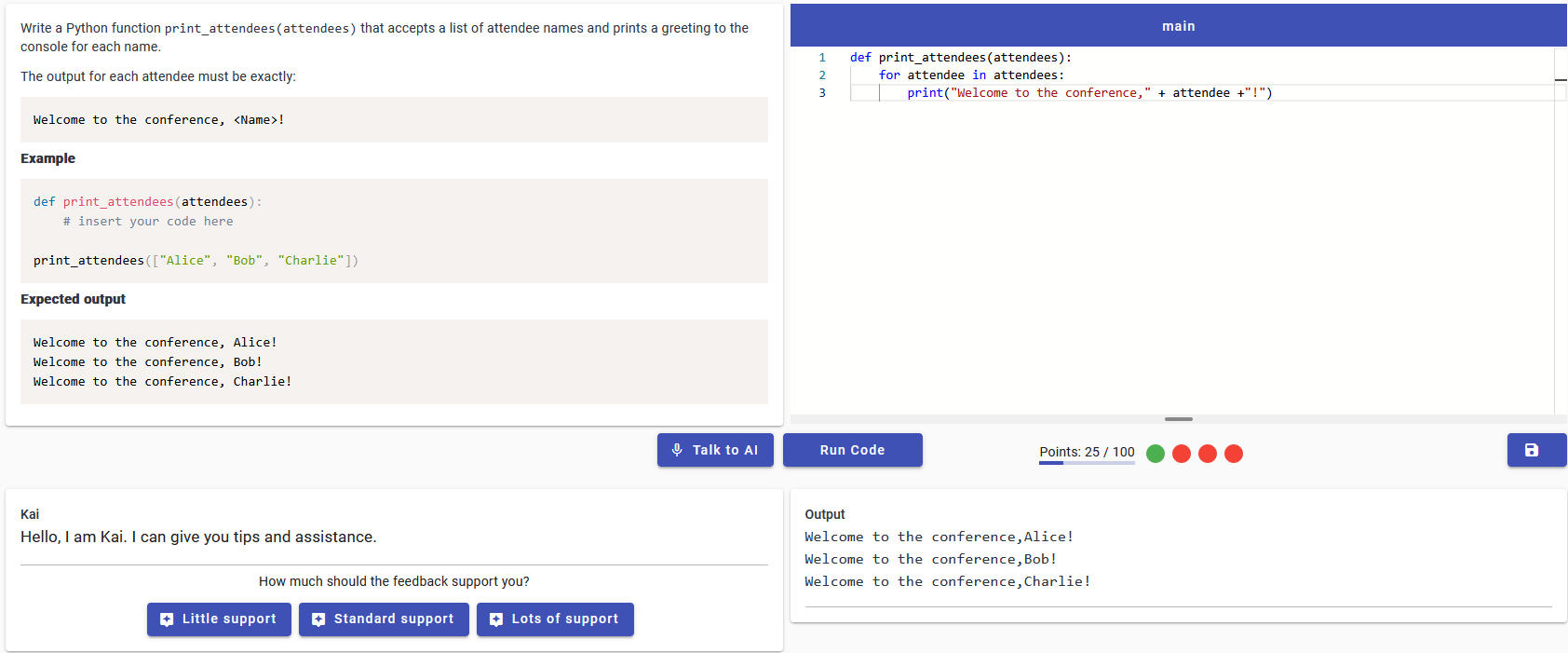}
    \caption{Tutor Kai User Interface}
    \label{fig: Tutor Kai User Interface}
\end{figure}
\section{Voice Generation in the Tutor Kai} \label{sec: Voice Generation in the Tutor Kai}
The Tutor Kai is a learning environment designed to support students in solving programming tasks. The standard interface presents students with a task description (left) next to a code editor (right, see \autoref{fig: Tutor Kai User Interface}). Students can execute their code within the environment, receiving immediate feedback through compiler output and the results from automated unit tests associated with the task. In addition, the Tutor Kai offers the option to generate textual feedback on demand via GenAI integration \cite{jacobs_evaluating_2024, jacobs.2024c, jacobs2025unlimited, jacobs.2025a}.

We integrated a real-time voice mode into the Tutor Kai.
The new feature was implemented using OpenAI's Realtime API, specifically the \emph{gpt-4o-realtime-preview-2024-10-01} model endpoint available at the time of the study (Dec. 2024), leveraging the available multimodal (voice) capabilities of OpenAI's GPT-4o model~\cite{openai_gpt-4o_2024}.
The API can handle streaming audio input and output, enabling low-latency, conversational interaction. Based on preliminary testing, we selected the \emph{sage} voice option for the voice output, as it offered the most natural and friendly-sounding German pronunciation among the available choices. Other than that, standard API defaults, including automatic voice activity detection (VAD) to manage turn-taking and a temperature setting of 0.6 were used.

To activate/deactivate the voice tutor, students can use a dedicated button below the task description (see \autoref{fig: Tutor Kai User Interface}). It provides visual feedback when the microphone is active (pulsating icon in red). While the voice mode is active, the system uses VAD to detect if the student speaks. The student's audio input is processed upon speech detection, and the AI generates an audio response with human-like delay (average of 320 milliseconds \cite{openai_gpt-4o_2024}).
A critical aspect of the implementation is the dynamic integration of task-specific context. When a student verbalizes input, the system automatically supplements the audio data with textual context before sending it to the Realtime API. This contextual information includes: (1) the student's current code from the editor, (2) the most recent compiler output (if any), and (3) the preceding turns of the current dialogue. A dialogue consists of all audio messages recorded during voice mode activation on a particular task.
The system prompt defines the AI's persona as a helpful tutor, sets constraints (e.g., forbidding direct solution revelation, encouraging brevity), and explicitly instructs the AI to reference the provided code and compiler output. Based on findings from our pilot study (see \autoref{sec: Pre Study Test}), the prompt directs the model to \say{describe code colloquially, as if in conversation}. The system prompt also includes the task description and the code skeleton that students get for each exercise.
By providing context, the GenAI voice tutor can be spoken to like a human tutor who sees the student's screen.
The system prompt and a demo recording of the GenAI voice tutor are available in our supplementary data repository \cite{supplementarydata.2025}.

\section{Pilot Study} \label{sec: Pre Study Test}
Before the main study, we conducted preliminary usability tests with the Voice Tutor. 
The goals were to gather user feedback on the voice interaction experience and identify potential technical or usability issues.
We recruited eight students (five male, three female) enrolled in a CS1 course at a German university. Each of them interacted with the voice-enabled tutor for about 20 minutes, working on 1-2 C++ exercises.
Their perceptions were gathered and evaluated via open-ended questions. They were generally positive and perceived as helpful. One student, however, perceived it as \say{unfamiliar and somewhat uncanny}, suggesting novelty effects or discomfort.
Moreover, students reported that the literal vocalization of code (e.g., \texttt{<}, \texttt{>}, and \texttt{\&\&}) was wrong and detrimental to comprehension, making it difficult to follow the AI's explanation or identify errors. Upon reviewing the system logs and audio recordings, we discovered that the textual logs provided by the OpenAI Realtime API for the AI's responses were not consistent with the audio output.
Examples of these issues are part of our supplementary data~\cite{supplementarydata.2025}. These findings informed the design and methodology of the main study. 
For example, the unreliability of the API's text logs necessitated a data analysis approach based entirely on the recorded audio for the main study, involving transcription and manual verification.
To mitigate the Voice Tutor's incorrect code pronunciation, we refined the system prompt to make sure code is described colloquially (see \autoref{sec: Voice Generation in the Tutor Kai}).

\section{Methodology}

\subsection{Data Collection} \label{subsec:DataCollection}

\subsubsection{Setting and Context}
The data were collected between Dec. 2024 and Jan. 2025 during eight regular computer science lessons (twice per week) of a ninth-grade class at a German grammar school. These lessons were part of the curricular unit \say{Introduction to the Python Programming Language}. The study was conducted in an authentic classroom environment where students engaged in typical learning activities, including potential interactions with peers and using external resources such as web search or other AI tools alongside the provided Tutor Kai. 
Audio input and output for the voice feature were facilitated through individual headsets (Logitech 960) supplied to the participating students. 

\subsubsection{Participants and Recruitment} \label{subsubsec:Participants}
The ninth-grade class comprised 15 students. Informed consent was obtained (following local regulations on research in schools) from the school administration, from the students' parents/legal guardians, and the students themselves. All consents were obtained for 11 out of 15 students. Subsequently, the GenAI Voice Tutor was enabled only for these 11 students. Participation was voluntary, and pseudonyms were used throughout data collection and analysis. Despite having consented, two students opted not to use the GenAI Voice Tutor during the study. So, the final participant group comprised nine students (six male, three female). Their ages ranged from 14 to 15 years (Mean (M) = 14.22, Standard deviation (SD) = 0.44).

\subsubsection{Procedure and Data Logging} \label{subsubsec:Procedure}
Data were gathered during regular computer science lessons as students worked on programming exercises within the Tutor Kai. These exercises were designed to reinforce the Python concepts being taught in the unit. 
While working on these tasks, participating students could activate or deactivate the GenAI Voice Tutor (as described in \autoref{sec: Voice Generation in the Tutor Kai}). Every student's voice input and corresponding GenAI voice response during an active session were automatically recorded as a separate audio file. These audio recordings were stored along with contextual information in a database, e.g., timestamp, the student's code state in the editor at the time of the request, the latest compiler output, and the preceding dialogue history (see Section~\ref{sec: Voice Generation in the Tutor Kai}). In the final lesson, students completed the German version \cite{seaborn_cross-cultural_2024} of the Partner Modelling Questionnaire (PMQ) \cite{Seaborn2025data,doyle_partner_2023} to capture their overall perceptions of the GenAI Voice Tutor as a dialogue partner. It allows for further data triangulation. 

\begin{table*}[thb]
  \centering
  \footnotesize
  \caption{Final coding scheme with categories and examples for student voice prompt analysis}
  \label{tab: Categories for student voice prompt analysis}
  \begin{tabular}{@{} 
      >{\raggedright\arraybackslash}p{0.18\textwidth}
      >{\raggedright\arraybackslash}p{0.33\textwidth}
      >{\raggedright\arraybackslash}p{0.43\textwidth}
    @{}}
    \toprule
    \textbf{Category}
      & \textbf{Definition}
      & \textbf{Anchor Examples} \\
    \midrule

    \multicolumn{2}{@{}p{0.40\textwidth}@{}}{
      \makecell[tl]{\emph{Other:}}
    }
      & \\
    \addlinespace[2pt]

    *Work alone first (WAF) (new)
      & Student does not want feedback yet.
      & \makecell[tl]{\say{Let me try it myself first.}, \say{I think I can manage on my own for now.}} \\

    \addlinespace[2pt]
    *Small talk (ST) (new)
      & Social interactions unrelated to the task description and computer science.
      & \makecell[tl]{\say{It’s okay. You don’t have to apologize.}\\
                     \say{See you next time.}} \\                    

    \addlinespace[2pt]
    Solution request (SR)
      & Prompt explicitly seeks a direct and complete solution.
      & \makecell[tl]{\say{Solve the task for me.}, \say{Can you tell me the solution?} (Hypothetical)}  \\

    \midrule

    \multicolumn{2}{@{}p{0.40\textwidth}@{}}{
      \makecell[tl]{\emph{Type of follow‐up interaction (based on \citet{scholl_analyzing_2024}):}}
    }
      & \\
    \addlinespace[2pt]

    Standalone query (STA)
      & Initial or independent prompt not referring to previous conversation.
      & \makecell[tl]{\say{Hey, I don’t understand what I’m supposed to do. Can you help me?}\\
                     \say{Is this correct?}} \\

    \addlinespace[2pt]
    Follow-up to a previous prompt (PRE)
      & Student references or builds on their own previous prompt.
      & \makecell[tl]{\say{Why am I still getting an error message?}\\
                     \say{Now correct?}} \\

    \addlinespace[2pt]
    Response to GenAI answer (RES)
      & Student directly responds to something in the GenAI’s reply.
      & \makecell[tl]{\say{What do you mean?}\\
                     \say{How do I do that?}} \\

    \addlinespace[2pt]
    Correction (COR)
      & Student attempts to correct or refine GPT’s output (content, tone, or speed).
      & \makecell[tl]{\say{Wait, wait. That’s a bit fast. I can’t keep up.}\\
                     \say{Yes, but it says it’s wrong.}} \\

    \midrule

    \multicolumn{3}{@{}p{0.60\textwidth}@{}}{
      \makecell[tl]{\emph{Issue or problem solving step adressed in follow-up interaction (based on \citet{scholl_analyzing_2024}):}}
    }
      \\
    \addlinespace[2pt]

    Debugging (DE)
      & Identifying and fixing errors in existing code.
      & \makecell[tl]{\say{What’s wrong with my loop?}, \say{And how do I fix the problem?}} \\

    \addlinespace[2pt]
    Problem Understanding (PU)
      & Seeking to understand the problem statement (e.g., task description).
      & \makecell[tl]{\say{Explain the task to me.}\\
                     \say{What exactly does the task want me to output?}} \\

    \addlinespace[2pt]
    Conceptual Understanding (CU)
      & Seeking explanation of programming concepts or theory.
      & \makecell[tl]{\say{I didn’t quite understand the difference between a variable and a parameter.}\\
                     \say{I still haven’t got it. Can you explain it as if I were a young child?}} \\

    \addlinespace[2pt]
    Syntax Style (SY)
      & Questions about code formatting or stylistic elements.
      & \makecell[tl]{\say{What do you mean? Quotation marks within quotation marks?}\\
                     \say{Does there need to be a space between the plus sign and the numbers?}} \\

    \addlinespace[2pt]
    *Pair Programming (PP) (new)
      & Asking for support during the first solution approach.
      & \makecell[tl]{\say{And now?}, \say{Okay, I’ve defined it. What’s next?}} \\

    \addlinespace[2pt]
    Other (OT)
      & Other utterances.
      & \makecell[tl]{\say{But it really doesn’t make any sense right now, sorry.}\\
                     \say{You don’t have to repeat yourself all the time, okay?}} \\

    \bottomrule
  \end{tabular}
\end{table*}

\subsection{Data Analysis} \label{subsec: Data Analysis}
Given the discrepancy between the Realtime API's text logs and the actual audio output identified during the pilot (see \autoref{sec: Pre Study Test}), we transcribed all audio dialogues using the open-weight \emph{whisper-large-v3-turbo} speech recognition model~\cite{radford_robust_2022} on a local machine. Next, we reviewed all transcripts against the original audio recordings to ensure accuracy. 
During the review of the transcripts, the data were cleaned. For example, unintended voice inputs triggered by background noise, non-verbal utterances (e.g., sighs, coughs), thinking aloud, or external conversations were removed. Thinking aloud was recognizable, e.g., via whispered repetitions of responses from the GenAI Voice Tutor or pronouncing individual code elements that the student added to their code.
Dialogues consisting \emph{exclusively} of such unintended voice inputs and their corresponding GenAI voice responses were discarded from the dataset before coding in MAXQDA \cite{kuckartz_analyzing_2019}. However, unintended voice inputs occurring within otherwise meaningful dialogues were kept but marked during the coding so we could distinguish them from intentional prompts.
All other voice inputs were classified as \emph{voice prompts}.

\subsubsection{Student Voice Prompt Analysis (RQ1)} \label{subsubsec: Student Voice Prompt Analysis (RQ1)}

Within the 608 analyzed student voice inputs, 205 were identified as \emph{unintended voice inputs}. The main cause (69 instances) was students thinking aloud (e.g., whispering to themselves), which was picked up by the microphone and triggered a response. 
Consequently, our analysis of student interaction patterns focuses on the 403 intended \emph{student voice prompts}. 
To address RQ1, we analyzed the content of the remaining intentional student voice prompts ($n=403$). The entire related dialog was used as a context unit, if needed. We reused the coding scheme developed by \citet{scholl_analyzing_2024} for deductively analyzing text-based ChatGPT interactions in introductory programming.
As our data contained new insights, some new categories were developed inductively (e.g., work alone first (WAF), small talk (ST), and pair programming (PP), see \autoref{tab: Categories for student voice prompt analysis}). 
Other categories from related work (code generation, runtime analysis, documentation, and test cases) were removed because they did not occur during coding.
The final coding scheme (see \autoref{tab: Categories for student voice prompt analysis}) was applied to each student voice prompt by one author in two steps.
First, it was assigned to exactly one \emph{Type of follow-up interaction} (e.g., STA, PRE, RES, COR, see \autoref{tab: Categories for student voice prompt analysis}) or an \emph{Other} category (e.g., WAF, ST, SR, see \autoref{tab: Categories for student voice prompt analysis}).
Second, prompts coded as follow-up interaction were further coded for one or more \emph{Issue or problem-solving step addressed in follow-up interaction} (e.g., DE, PU, CU, SY, PP, OT, see \autoref{tab: Categories for student voice prompt analysis}).
The entire material was coded twice by the same author, two weeks apart.

\subsubsection{GenAI Voice Response Analysis (RQ2)} \label{subsubsec: GenAI Voice Response Analysis (RQ2)}
To address RQ2, we analyzed all GenAI voice responses ($n=602$) including those triggered by unintended inputs. We retained these as the surrounding dialogue indicated active student engagement immediately before and after the unintended input, i.e., students likely heard and potentially processed these responses.
For the coding process, we used a student's entire dialog and related information (task description, current code, and compiler output) as a context unit. First, we categorized each response's overall type, distinguishing between Interrupted responses (IR-responses interrupted by student voice inputs), Small talk (ST-responses unrelated to the programming task and computer science), Feedback (FB), and Other (OT).
Second, we performed a detailed analysis on the responses categorized as feedback (FB, $n=416$) using human intelligence. This involved:
\begin{enumerate}[leftmargin=*]
    \item \emph{Feedback Type}: We classified the feedback types based on the typology for programming by \citet{keuning_systematic_2019} and \citet{narciss2008feedback} (e.g., knowledge of result (KR), etc.). Multiple types could be assigned to a GenAI feedback.
    \item \emph{Correctness}: Each feedback was assessed for correctness (binary: Feedback Correct - FC vs. Feedback Not Correct - FNC).
    \item \emph{Issues}: We coded issues related to the voice delivery and content generation, namely Language Incorrect (LAI - responses that were nonsense or poorly vocalized code) and Repetition (REP - instances where the AI repeated itself multiple times).
\end{enumerate}

\subsubsection{Partner Modelling Questionnaire Analysis (RQ3)} \label{ssubsec:PMQAnalysis}
To address RQ3 and triangulate our findings, we analyzed the responses from the nine participating students to the German version \cite{Seaborn2025data,seaborn_cross-cultural_2024} of the Partner Modelling Questionnaire (PMQ) \cite{doyle_partner_2023}. 
Following the procedure described by \citet{doyle_partner_2023}, we calculated descriptive statistics (mean and standard deviation) for the overall PMQ score and its constituent subscales (aggregated for the area Competence and Dependability, Human-Likeness, and Communicative Flexibility) as demanded for this particular instrument~\cite{doyle_partner_2023}.

\section{Results} \label{sec: Results}

Over the four-week data collection period, the nine participating students generated 3172 audio files (student requests and AI responses combined) using the GenAI Voice Mode feature of the Tutor Kai. 
Following the data cleaning procedure described in section \ref{subsec: Data Analysis}, the final dataset comprised 1210 audio recordings (608 student voice inputs and 602 GenAI voice responses), constituting 50 distinct dialogues.
The remaining 1962 recordings were discarded, primarily because they belonged to sessions where the automatic voice activity detection (VAD) remained active for extended periods without intentional interactions (often due to students forgetting to deactivate the feature), resulting in dialogues composed exclusively of \emph{unintended voice inputs} and responses.
Dialogues consisted of an average of $M=24.2$ messages ($SD=17.17$, $Min=4$, $Max=88$). The transcribed GenAI voice responses averaged $M=181.15$ characters in length ($SD=178.21$, $Min=2$, $Max=2324$) and transcribed student voice inputs averaged $M=27.47$ characters ($SD=27.37$, $Min=2$, $Max=208$).

\subsection{RQ1: Student Prompts and Interactions} \label{subsec:RQ1_Results}

To answer RQ1, we focused on the 608 student voice inputs. 205 of them were identified as unintended inputs (see section~\ref{subsubsec: Student Voice Prompt Analysis (RQ1)}). 
The 403 intentional student voice prompts were analyzed via the coding scheme in \autoref{tab: Categories for student voice prompt analysis}. The results reveal how students interacted with the GenAI Voice Tutor (see \autoref{tab:quantification-students-voice-prompts}).
In 18.9\% of cases (76 out of 403), prompts constituted \emph{Small talk} (ST). This typically occurred as a greeting at the beginning of a session, sometimes to confirm audio transmission (e.g., \say{Hello, can you hear me?}). In 4.5\% (18) instances, students explicitly indicated a desire to \emph{Work alone first} (WAF), often declining an AI response triggered by an unintended voice input while they were working. Explicit full \emph{Solution requests} (SR) were absent (0.0\%).

\begin{table}[htb]
  \centering
  \footnotesize
  \caption{Quantification of students voice prompts}
  \label{tab:quantification-students-voice-prompts}
  \begin{tabular}{@{}l r r r r r r r@{}}
    \toprule
    \multicolumn{1}{@{}l}{Prompt Category} 
      & \multicolumn{1}{c}{Count (\%)} 
      & \multicolumn{1}{c}{} 
      & \multicolumn{1}{c}{} 
      & \multicolumn{1}{c}{} 
      & \multicolumn{1}{c}{} 
      & \multicolumn{1}{c}{} 
      & \multicolumn{1}{c@{}}{} \\
    \midrule
    Work alone first (WAF)      & 18 (4.47\%)  &    &    &    &    &    &    \\
    Small talk (ST)             & 76 (18.86\%) &    &    &    &    &    &    \\
    Solution request (SR)       & 0 (0.00\%)   &    &    &    &    &    &    \\
    \midrule
    Type of interaction & & DE & PU & CU & SY & PP & OT\\ 
    \midrule
    Standalone query (STA)      & 102 (25.31\%) & 72 &  9 &  5 & 1 &  14 &  3 \\
    Follow-up to previous (PRE) &  52 (12.90\%) & 34 & 2 & 4 & 2 & 10 & 0    \\
    Response to GenAI (RES)     & 121 (30.02\%)& 46 & 13 & 16 & 7 & 35 & 13  \\
    Correction (COR)            &  34 (8.44\%) & 22 & 1 & 0 & 0 & 3 & 8    \\
    \bottomrule
  \end{tabular}
\end{table}
\noindent
Analyzing the \emph{Type of Interaction} showed that students usually replied directly to the GenAI voice answer (RES, 30.0\%, 121 prompts). \emph{Standalone queries} were also common as initial requests (STA, 25.3\%, 102 prompts). These were often followed up in other prompts (PRE, 12.9\%, 52 prompts). Students explicitly attempted to \emph{Correct} the AI (COR) in 8.4\% (34 prompts) of cases.

We analyzed the (one or multiple) \emph{Issues or problem solving steps} addressed within 309 interactions (coded as STA, PRE, RES, COR types). 
\emph{Debugging} (DE) was the most prominent focus, involved in 56.3\% (174 out of 309) of these interaction prompts. In these instances, students typically sought help with existing solution attempts. 
The second most frequent focus was \emph{Pair Programming} (PP), involved in 20.1\% (62 out of 309) of interaction prompts, where students used the GenAI Voice Tutor to develop a solution approach incrementally. 
Queries about \emph{Conceptual Understanding} (CU) and \emph{Problem Understanding} (PU) were less frequent but present (both 8.1\%, 25 out of 309 prompts). 
Questions regarding \emph{Syntax/Style} (SY) were rare (3.2\%, 10 out of 309 prompts).

\subsection{RQ2: GenAI Voice Responses} \label{subsec: RQ2 Results}

We analyzed all 602 GenAI voice responses to characterize the output (see \autoref{tab: Quantification of GenAI voice answers}). 
The majority of responses were classified as \emph{Feedback} (FB, 69.1\%, 416 responses). 
\emph{Small talk} (ST), often responding to the students' greeting (e.g., at the beginning of a conversation), was also frequent (23.3\%, 140 responses). 
Responses were \emph{Interrupted} (IR) by the student or system in 7.0\% (42 responses) of cases, and a small fraction were classified as \emph{Other} (OT, 0.7\%, 4 responses).

The 416 GenAI voice responses classified as \emph{Feedback} (FB) were analyzed further w.r.t. their feedback types, correctness, and quality issues (\autoref{tab: Feedback Analysis}). The most frequent type \cite{keuning_systematic_2019,narciss2008feedback} was \emph{Knowledge on how to proceed} (KH), present in 70.4\% (293) of feedback responses, often providing next-step hints. \emph{Knowledge about mistakes} (KM) was also common (36.3\%, 151 responses). 
Other feedback types occurred less frequently (see \autoref{tab: Feedback Analysis} for full details). 
It should be noted that a single GenAI voice response may have contained multiple feedback types.
\begin{table}[htb]
  \centering
  \footnotesize
  \caption{Quantification of GenAI voice answers}
  \label{tab: Quantification of GenAI voice answers}
  \begin{tabular}{@{}l r@{}}
    \toprule
    GenAI voice answers category & Count (\%)\\ 
    \midrule
    Interrupted (IR)  & 42 (6.98\%)\\
    Small talk (ST) & 140 (23.26\%)\\
    Feedback (FB) & 416 (69.10\%)\\
    Other (OT) & 4 (0,66\%)  \\
    \bottomrule
  \end{tabular}
\end{table}
\begin{table}[ht]
  \centering
  \footnotesize
  \caption{Feedback Type Analysis}
  \label{tab: Feedback Analysis}
  \begin{tabular}{@{}l r r r r@{}}
    \toprule
    Feedback-Type & Count (\%) & FC & FNC & LAI \\ 
    \midrule
    Knowledge of result (KR)                & 52 (12.50\%) & 23 & 29 & 1 \\
    Knowledge of correct result (KCR)       & 22 (5.29\%) & 6 & 16 &  12 \\
    Knowledge of performance (KP)           & 1 (0.24\%) & 1 & 0 & 0 \\
    Knowledge about task constraints (KTC)  & 16 (3.85\%) & 16 & 0 & 1 \\
    Knowledge about concepts (KC)           & 19 (4.57\%) & 17 & 2 & 0 \\
    Knowledge about mistakes (KM)           & 151 (36.30\%) & 111 & 40 & 25 \\
    Knowledge on how to proceed (KH)        & 293 (70.43\%) & 219 & 74 & 53 \\
    Knowledge about meta-cognition (KMC)    & 5 (1.20\%) & 5 & 0 & 0 \\
    \bottomrule
  \end{tabular}
\end{table}
\noindent
Concerning \emph{Correctness}, 71.4\% (297 out of 416) of the feedback instances were judged as correct (FC), while 28.6\% (119) were assessed as not correct (FNC). The correctness greatly varied across feedback types. While feedback on task constraints (KTC) and concepts (KC) was mainly correct, feedback providing the correct result (KCR) was incorrect in 72.7\% (16 out of 22) of cases. 
Similarly, feedback on how to proceed (KH) was incorrect 25.3\% (74 out of 293) of the time, and feedback about mistakes (KM) was incorrect 26.5\% (40 out of 151) of the time.

Regarding \emph{Quality Issues}, we specifically tracked linguistic problems (LAI) and repetition (REP). Incorrect language or nonsense output (LAI) occurred in 69 feedback instances. As indicated in \autoref{tab: Feedback Analysis}, LAI frequently coincided with incorrect feedback (FNC), particularly for KCR (12 LAI cases) and KH (53 LAI cases). 
This aligns with the pilot study's findings (\autoref{sec: Pre Study Test}) that the GenAI Voice Tutor struggled with verbalizing specific code elements, despite the prompt instruction to avoid reading code aloud. 
Excessive repetition (REP) occurred in 7 instances, where the GenAI Voice Tutor repeated the same phrase or sentence multiple times, sometimes for extended durations (up to 3 minutes).

\subsection{RQ3: Student Perceptions of the Voice Tutor} \label{subsec:RQ3_Results}

The Partner Modelling Questionnaire (PMQ) was used to assess students' overall perceptions of the GenAI Voice Tutor as a dialogue partner ($n=9$).
The overall mean score on the PMQ was moderate (Mean (M) = 4.23, Standard Deviation (SD) = 1.24) on the 7-point scale. Students rated the GenAI Voice Tutor relatively high for the area \emph{Competence and Dependability} (M = 5.65, SD = 0.21), suggesting they perceived it as generally capable and reliable for the tasks. However, ratings for \emph{Human-Likeness} (M = 3.41, SD = 0.52) and \emph{Communicative Flexibility} (M = 3.63, SD = 0.61) were lower and in the neutral area of the scales.

\section{Discussion}

The 9th-graders in our study engaged in dialogues with the GenAI Voice Tutor, evident via the RES (Response to GenAI) and PRE (follow-up) voice prompts, as well as the average of 24.2 messages per dialogue.
The GenAI Voice Tutor's context-awareness (task description, current program code, and compiler output) helped support students' very brief voice prompts (e.g., \say{Correct?}), also observed by \citet{jaber_cooking_2024}. We consider this advantageous for novice students already challenged by learning a programming language, such as Python. 

The GenAI Voice Tutor also responded to student requests (cf. \cite{scholl2025students}): KH (Knowledge on how to proceed) and KM (Knowledge about mistakes) were the most frequent feedback categories, aligning with students' most frequent voice interactions and prompts (DE and PP). Similarly, KC and KTC feedback types as part of the generated output matched CU and PU in student voice prompts. Social interactions unrelated to the programming task, such as greetings or thanking the Tutor Kai (classified as Small Talk, ST), comprised 18.9\% of students' voice prompts. The GenAI Tutor often mirrored these exchanges, with 23.3\% of its voice responses also being small talk. 
While this frequency of ST may suggest that students anthropomorphize the GenAI Voice Tutor, the triangulation with the Partner Modelling Questionnaire (PMQ) ratings offers another perspective: Students assigned neutral scores for \emph{Human-Likeness} (M=3.41 on a 7-point scale) and \emph{Communicative Flexibility} (M=3.63 on a 7-point scale).
The linguistic struggles of the GenAI Voice Tutor when verbalizing code (LAI) likely played a role here. Furthermore, students may have also compared the GenAI's voice output not just to GenAI's textual output, but to the richness and nuance of human communication. Hence, the expectation of human-likeness may have been quite high.

A critical observation concerns the correctness of the GenAI Voice Tutor's responses, as 28.6\% of all feedback instances were incorrect (FNC). Its difficulty in verbalizing programming code was the most influential factor in the poor output quality (LAI). Despite the inclusion of \say{describe code colloquially} into the system prompt (see section \ref{sec: Voice Generation in the Tutor Kai}), the tool frequently attempted to read symbols and structures literally, rendering the output confusing and often incorrect.
Due to the high error rate in the generated feedback, students may fall into drifting patterns \cite{prather_its_2023}, being led down misleading paths by incorrect information in GenAI voice answers. Considering diverse learners and the potential of voice assistance for them, these results are somewhat disappointing.

Despite these flaws in the feedback (high FNC rate and LAI issues), the triangulation with the PMQ results shows high ratings for \emph{Competence and Dependability} (M=5.65/7). Yet, due to students' limited domain knowledge, novices may not always recognize incorrect suggestions.

\section{Threats to Validity}
Due to the study's authentic classroom setting, we could not fully control all variables, e.g., students' interactions with peers or their use of external resources like web searches or other AI tools. 
The nature of the study is qualitative and explorative. With a sample size of 9 students, we have reached saturation~\cite{boddy2016sample} within the material. This was due to the 1210 audio messages we analyzed in sum. However, we do not claim to provide generalizations or transferable conclusions about 9th-graders' interaction patterns and perceptions on GenAI voice outputs. 
It should further be noted that the Partner Modelling Questionnaire (RQ3) only captures student perceptions at a single point in time, which may have been influenced by novelty effects or their most recent interactions.

\section{Conclusions}
In this study, we evaluated how novice programmers engage with a multimodal (voice) Generative AI tutor in an authentic secondary school classroom. By analyzing 1210 student-AI audio messages from nine ninth-graders learning Python, and their perceptions via the Partner Modelling Questionnaire, we reached the following conclusions: (1) Novices primarily used the voice tutor for debugging (56.3\% of voice interaction prompts) and pair programming-like scenario (20.1\%), engaging in dialogues (M=24.2 messages) with the GenAI Voice Tutor without seeking direct solutions. (2) The GenAI's voice response, predominantly offering guidance (70.4\% of feedback instances) on how to proceed and error explanation (36.3\%), was unreliable: 28.6\% of its 416 feedback instances were incorrect. A key issue was its persistent poor verbalization of programming code, making it too unreliable for novices, let alone for learners with diverse needs and skills.

While the GenAI Voice Tutor offers potential for natural, context-aware interactions that could benefit accessibility at some point, its current output renders it problematic. Without improvements, the GenAI Voice Tutor risks even more challenges for learners requiring additional support.  
Future work should enhance GenAI's capability to \say{speak (about) code fluently}. Once a baseline of reliability and safety of the voice mode can be assured, further studies with diverse learners, particularly those with disabilities, are essential and encouraged.

\bibliographystyle{ACM-Reference-Format}
\balance
\bibliography{bib}


\begin{thebibliography}{47}


\ifx \showCODEN    \undefined \def \showCODEN     #1{\unskip}     \fi
\ifx \showISBNx    \undefined \def \showISBNx     #1{\unskip}     \fi
\ifx \showISBNxiii \undefined \def \showISBNxiii  #1{\unskip}     \fi
\ifx \showISSN     \undefined \def \showISSN      #1{\unskip}     \fi
\ifx \showLCCN     \undefined \def \showLCCN      #1{\unskip}     \fi
\ifx \shownote     \undefined \def \shownote      #1{#1}          \fi
\ifx \showarticletitle \undefined \def \showarticletitle #1{#1}   \fi
\ifx \showURL      \undefined \def \showURL       {\relax}        \fi
\providecommand\bibfield[2]{#2}
\providecommand\bibinfo[2]{#2}
\providecommand\natexlab[1]{#1}
\providecommand\showeprint[2][]{arXiv:#2}

\bibitem[Alfageeh et~al\mbox{.}(2025)]%
        {alfageeh_prompts_2025}
\bibfield{author}{\bibinfo{person}{Ali Alfageeh}, \bibinfo{person}{Sadegh AlMahdi~Kazemi Zarkouei}, \bibinfo{person}{Daye Nam}, \bibinfo{person}{Daniel Prol}, \bibinfo{person}{Matin Amoozadeh}, \bibinfo{person}{Souti Chattopadhyay}, \bibinfo{person}{James Prather}, \bibinfo{person}{Paul Denny}, \bibinfo{person}{Juho Leinonen}, \bibinfo{person}{Michael Hilton}, \bibinfo{person}{Sruti~Srinivasa Ragavan}, {and} \bibinfo{person}{Mohammad~Amin Alipour}.} \bibinfo{year}{2025}\natexlab{}.
\newblock \bibinfo{title}{From {Prompts} to {Propositions}: {A} {Logic}-{Based} {Lens} on {Student}-{LLM} {Interactions}}.
\newblock
\href{https://doi.org/10.48550/arXiv.2504.18691}{doi:\nolinkurl{10.48550/arXiv.2504.18691}}


\bibitem[Alshaigy and Grande(2024)]%
        {alshaigy2024forgotten}
\bibfield{author}{\bibinfo{person}{Bedour Alshaigy} {and} \bibinfo{person}{Virginia Grande}.} \bibinfo{year}{2024}\natexlab{}.
\newblock \showarticletitle{Forgotten Again: Addressing Accessibility Challenges of Generative AI Tools for People with Disabilities}. In \bibinfo{booktitle}{\emph{Adjunct Proceedings of the 2024 Nordic Conference on Human-Computer Interaction}}. \bibinfo{publisher}{ACM}, \bibinfo{address}{New York}.
\newblock
\showISBNx{9798400709654}
\href{https://doi.org/10.1145/3677045.3685493}{doi:\nolinkurl{10.1145/3677045.3685493}}


\bibitem[Amoozadeh et~al\mbox{.}(2024)]%
        {amoozadeh_student-ai_2024}
\bibfield{author}{\bibinfo{person}{Matin Amoozadeh}, \bibinfo{person}{Daye Nam}, \bibinfo{person}{Daniel Prol}, \bibinfo{person}{Ali Alfageeh}, \bibinfo{person}{James Prather}, \bibinfo{person}{Michael Hilton}, \bibinfo{person}{Sruti Srinivasa~Ragavan}, {and} \bibinfo{person}{Amin Alipour}.} \bibinfo{year}{2024}\natexlab{}.
\newblock \showarticletitle{Student-AI Interaction: A Case Study of CS1 students}. In \bibinfo{booktitle}{\emph{Proceedings of the 24th Koli Calling International Conference on Computing Education Research}} \emph{(\bibinfo{series}{Koli Calling '24})}. \bibinfo{publisher}{ACM}, \bibinfo{address}{New York}, Article \bibinfo{articleno}{13}.
\newblock
\showISBNx{9798400710384}
\href{https://doi.org/10.1145/3699538.3699567}{doi:\nolinkurl{10.1145/3699538.3699567}}


\bibitem[Apiola et~al\mbox{.}(2024)]%
        {apiola2024first}
\bibfield{author}{\bibinfo{person}{Mikko Apiola}, \bibinfo{person}{Henriikka Vartiainen}, {and} \bibinfo{person}{Matti Tedre}.} \bibinfo{year}{2024}\natexlab{}.
\newblock \showarticletitle{First Year CS Students Exploring And Identifying Biases and Social Injustices in Text-to-Image Generative AI}. In \bibinfo{booktitle}{\emph{Proceedings of the 2024 on Innovation and Technology in Computer Science Education V. 1}}. \bibinfo{publisher}{ACM}, \bibinfo{address}{New York}, \bibinfo{pages}{485–491}.
\newblock
\showISBNx{9798400706004}
\href{https://doi.org/10.1145/3649217.3653596}{doi:\nolinkurl{10.1145/3649217.3653596}}


\bibitem[Barke et~al\mbox{.}(2023)]%
        {barke_grounded_2023}
\bibfield{author}{\bibinfo{person}{Shraddha Barke}, \bibinfo{person}{Michael~B. James}, {and} \bibinfo{person}{Nadia Polikarpova}.} \bibinfo{year}{2023}\natexlab{}.
\newblock \showarticletitle{Grounded {Copilot}: {How} {Programmers} {Interact} with {Code}-{Generating} {Models}}.
\newblock \bibinfo{journal}{\emph{Proceedings of the ACM on Programming Languages}}  \bibinfo{volume}{7} (\bibinfo{year}{2023}), \bibinfo{pages}{85--111}.
\newblock
\showISSN{2475-1421}
\href{https://doi.org/10.1145/3586030}{doi:\nolinkurl{10.1145/3586030}}


\bibitem[Becker and Fitzpatrick(2019)]%
        {becker20219what}
\bibfield{author}{\bibinfo{person}{Brett~A. Becker} {and} \bibinfo{person}{Thomas Fitzpatrick}.} \bibinfo{year}{2019}\natexlab{}.
\newblock \showarticletitle{What Do CS1 Syllabi Reveal About Our Expectations of Introductory Programming Students?}. In \bibinfo{booktitle}{\emph{Proceedings of the 50th ACM Technical Symposium on Computer Science Education}} \emph{(\bibinfo{series}{SIGCSE '19})}. \bibinfo{publisher}{ACM}, \bibinfo{address}{New York}, \bibinfo{pages}{1011–1017}.
\newblock
\showISBNx{9781450358903}
\href{https://doi.org/10.1145/3287324.3287485}{doi:\nolinkurl{10.1145/3287324.3287485}}


\bibitem[Bender et~al\mbox{.}(2021)]%
        {bender2021onthedangers}
\bibfield{author}{\bibinfo{person}{Emily~M. Bender}, \bibinfo{person}{Timnit Gebru}, \bibinfo{person}{Angelina McMillan-Major}, {and} \bibinfo{person}{Shmargaret Shmitchell}.} \bibinfo{year}{2021}\natexlab{}.
\newblock \showarticletitle{On the Dangers of Stochastic Parrots: Can Language Models Be Too Big?}. In \bibinfo{booktitle}{\emph{Proceedings of the 2021 ACM Conference on Fairness, Accountability, and Transparency}}. \bibinfo{publisher}{ACM}, \bibinfo{address}{New York}, \bibinfo{pages}{610–623}.
\newblock
\showISBNx{9781450383097}
\href{https://doi.org/10.1145/3442188.3445922}{doi:\nolinkurl{10.1145/3442188.3445922}}


\bibitem[Boddy(2016)]%
        {boddy2016sample}
\bibfield{author}{\bibinfo{person}{Clive~Roland Boddy}.} \bibinfo{year}{2016}\natexlab{}.
\newblock \showarticletitle{Sample size for qualitative research}.
\newblock \bibinfo{journal}{\emph{Qualitative market research: An international journal}} \bibinfo{volume}{19}, \bibinfo{number}{4} (\bibinfo{year}{2016}), \bibinfo{pages}{426--432}.
\newblock


\bibitem[Butler and Starkey(2024)]%
        {butler_ok_2024}
\bibfield{author}{\bibinfo{person}{Laura Butler} {and} \bibinfo{person}{Louise Starkey}.} \bibinfo{year}{2024}\natexlab{}.
\newblock \showarticletitle{{OK} {Google}, help me learn: an exploratory study of voice-activated artificial intelligence in the classroom}.
\newblock \bibinfo{journal}{\emph{Technology, Pedagogy and Education}} \bibinfo{volume}{33}, \bibinfo{number}{2} (\bibinfo{year}{2024}), \bibinfo{pages}{135--148}.
\newblock
\href{https://doi.org/10.1080/1475939X.2024.2311779}{doi:\nolinkurl{10.1080/1475939X.2024.2311779}}


\bibitem[Doyle et~al\mbox{.}(2023)]%
        {doyle_partner_2023}
\bibfield{author}{\bibinfo{person}{Philip~R. Doyle}, \bibinfo{person}{Iona Gessinger}, \bibinfo{person}{Justin Edwards}, \bibinfo{person}{Leigh Clark}, \bibinfo{person}{Odile Dumbleton}, \bibinfo{person}{Diego Garaialde}, \bibinfo{person}{Daniel Rough}, \bibinfo{person}{Anna Bleakley}, \bibinfo{person}{Holly~P. Branigan}, {and} \bibinfo{person}{Benjamin~R. Cowan}.} \bibinfo{year}{2023}\natexlab{}.
\newblock \bibinfo{title}{The {Partner} {Modelling} {Questionnaire}: {A} validated self-report measure of perceptions toward machines as dialogue partners}.
\newblock
\href{https://doi.org/10.48550/arXiv.2308.07164}{doi:\nolinkurl{10.48550/arXiv.2308.07164}}


\bibitem[Essel et~al\mbox{.}(2025)]%
        {essel_exploring_2025}
\bibfield{author}{\bibinfo{person}{Harry~Barton Essel}, \bibinfo{person}{Dimitrios Vlachopoulos}, \bibinfo{person}{Henry Nunoo-Mensah}, {and} \bibinfo{person}{John~Opuni Amankwa}.} \bibinfo{year}{2025}\natexlab{}.
\newblock \showarticletitle{Exploring the impact of {VoiceBots} on multimedia programming education among {Ghanaian} university students}.
\newblock \bibinfo{journal}{\emph{British Journal of Educational Technology}} \bibinfo{volume}{56}, \bibinfo{number}{1} (\bibinfo{year}{2025}), \bibinfo{pages}{276--295}.
\newblock
\showISSN{1467-8535}
\href{https://doi.org/10.1111/bjet.13504}{doi:\nolinkurl{10.1111/bjet.13504}}


\bibitem[Hattie and Timperley(2007)]%
        {hattietimperley2007}
\bibfield{author}{\bibinfo{person}{John Hattie} {and} \bibinfo{person}{Helen Timperley}.} \bibinfo{year}{2007}\natexlab{}.
\newblock \showarticletitle{The Power of Feedback}.
\newblock \bibinfo{journal}{\emph{Review of Educational Research}} \bibinfo{volume}{77}, \bibinfo{number}{1} (\bibinfo{year}{2007}), \bibinfo{pages}{81--112}.
\newblock
\href{https://doi.org/10.3102/003465430298487}{doi:\nolinkurl{10.3102/003465430298487}}


\bibitem[Jaber et~al\mbox{.}(2024)]%
        {jaber_cooking_2024}
\bibfield{author}{\bibinfo{person}{Razan Jaber}, \bibinfo{person}{Sabrina Zhong}, \bibinfo{person}{Sanna Kuoppamäki}, \bibinfo{person}{Aida Hosseini}, \bibinfo{person}{Iona Gessinger}, \bibinfo{person}{Duncan~P Brumby}, \bibinfo{person}{Benjamin~R. Cowan}, {and} \bibinfo{person}{Donald Mcmillan}.} \bibinfo{year}{2024}\natexlab{}.
\newblock \showarticletitle{Cooking {With} {Agents}: {Designing} {Context}-aware {Voice} {Interaction}}. In \bibinfo{booktitle}{\emph{Proceedings of the 2024 {CHI} {Conference} on {Human} {Factors} in {Computing} {Systems}}} \emph{(\bibinfo{series}{{CHI} '24})}. \bibinfo{publisher}{ACM}, \bibinfo{address}{New York}, \bibinfo{pages}{1--13}.
\newblock
\showISBNx{979-8-4007-0330-0}
\href{https://doi.org/10.1145/3613904.3642183}{doi:\nolinkurl{10.1145/3613904.3642183}}


\bibitem[Jacobs(2025)]%
        {supplementarydata.2025}
\bibfield{author}{\bibinfo{person}{Sven Jacobs}.} \bibinfo{year}{2025}\natexlab{}.
\newblock \bibinfo{title}{Data: GenAI Voice Mode in Programming Education}.
\newblock
\href{https://doi.org/10.17605/OSF.IO/BAM9Y}{doi:\nolinkurl{10.17605/OSF.IO/BAM9Y}}


\bibitem[Jacobs and Jaschke(2024a)]%
        {jacobs_evaluating_2024}
\bibfield{author}{\bibinfo{person}{Sven Jacobs} {and} \bibinfo{person}{Steffen Jaschke}.} \bibinfo{year}{2024}\natexlab{a}.
\newblock \showarticletitle{Evaluating the {Application} of {Large} {Language} {Models} to {Generate} {Feedback} in {Programming} {Education}}. In \bibinfo{booktitle}{\emph{2024 IEEE Global Engineering Education Conference (EDUCON)}}. \bibinfo{publisher}{IEEE}.
\newblock
\href{https://doi.org/10.1109/EDUCON60312.2024.10578838}{doi:\nolinkurl{10.1109/EDUCON60312.2024.10578838}}


\bibitem[Jacobs and Jaschke(2024b)]%
        {jacobs.2024c}
\bibfield{author}{\bibinfo{person}{Sven Jacobs} {and} \bibinfo{person}{Steffen Jaschke}.} \bibinfo{year}{2024}\natexlab{b}.
\newblock \showarticletitle{Leveraging {Lecture} {Content} for {Improved} {Feedback}: {Explorations} with {GPT}-4 and {Retrieval} {Augmented} {Generation}}. In \bibinfo{booktitle}{\emph{2024 36th {International} {Conference} on {Software} {Engineering} {Education} and {Training} ({CSEE}\&{T})}}. \bibinfo{pages}{1--5}.
\newblock
\href{https://doi.org/10.1109/CSEET62301.2024.10663001}{doi:\nolinkurl{10.1109/CSEET62301.2024.10663001}}


\bibitem[Jacobs et~al\mbox{.}(2025a)]%
        {jacobs.2025a}
\bibfield{author}{\bibinfo{person}{Sven Jacobs}, \bibinfo{person}{Maurice Kempf}, {and} \bibinfo{person}{Natalie Kiesler}.} \bibinfo{year}{2025}\natexlab{a}.
\newblock \showarticletitle{That's {Not} the {Feedback} {I} {Need}! - {Student} {Engagement} with {GenAI} {Feedback} in the {Tutor} {Kai}}. In \bibinfo{booktitle}{\emph{Proceedings of the 2025 {Conference} on {UK} and {Ireland} {Computing} {Education} {Research}}} \emph{(\bibinfo{series}{{UKICER} '25})}. \bibinfo{publisher}{ACM}, \bibinfo{address}{New York}, \bibinfo{pages}{1--7}.
\newblock
\showISBNx{979-8-4007-2078-9}
\href{https://doi.org/10.1145/3754508.3754512}{doi:\nolinkurl{10.1145/3754508.3754512}}


\bibitem[Jacobs et~al\mbox{.}(2025b)]%
        {jacobs2025unlimited}
\bibfield{author}{\bibinfo{person}{Sven Jacobs}, \bibinfo{person}{Henning Peters}, \bibinfo{person}{Steffen Jaschke}, {and} \bibinfo{person}{Natalie Kiesler}.} \bibinfo{year}{2025}\natexlab{b}.
\newblock \showarticletitle{Unlimited Practice Opportunities: Automated Generation of Comprehensive, Personalized Programming Tasks}. In \bibinfo{booktitle}{\emph{Proceedings of the 30th ACM Conference on Innovation and Technology in Computer Science Education V. 1}} \emph{(\bibinfo{series}{ITiCSE 2025})}. \bibinfo{publisher}{ACM}, \bibinfo{address}{New York}, \bibinfo{pages}{319–325}.
\newblock
\showISBNx{9798400715679}
\href{https://doi.org/10.1145/3724363.3729089}{doi:\nolinkurl{10.1145/3724363.3729089}}


\bibitem[Jošt et~al\mbox{.}(2024)]%
        {jost2024theimpact}
\bibfield{author}{\bibinfo{person}{Gregor Jošt}, \bibinfo{person}{Viktor Taneski}, {and} \bibinfo{person}{Sašo Karakatič}.} \bibinfo{year}{2024}\natexlab{}.
\newblock \showarticletitle{The Impact of Large Language Models on Programming Education and Student Learning Outcomes}.
\newblock \bibinfo{journal}{\emph{Applied Sciences}} \bibinfo{volume}{14}, \bibinfo{number}{10} (\bibinfo{year}{2024}).
\newblock
\showISSN{2076-3417}
\href{https://doi.org/10.3390/app14104115}{doi:\nolinkurl{10.3390/app14104115}}


\bibitem[Kazemitabaar et~al\mbox{.}(2024)]%
        {kazemitabaar_codeaid_2024}
\bibfield{author}{\bibinfo{person}{Majeed Kazemitabaar}, \bibinfo{person}{Runlong Ye}, \bibinfo{person}{Xiaoning Wang}, \bibinfo{person}{Austin~Zachary Henley}, \bibinfo{person}{Paul Denny}, \bibinfo{person}{Michelle Craig}, {and} \bibinfo{person}{Tovi Grossman}.} \bibinfo{year}{2024}\natexlab{}.
\newblock \showarticletitle{CodeAid: Evaluating a Classroom Deployment of an LLM-based Programming Assistant that Balances Student and Educator Needs}. In \bibinfo{booktitle}{\emph{Proceedings of the 2024 CHI Conference on Human Factors in Computing Systems}} \emph{(\bibinfo{series}{CHI '24})}. \bibinfo{publisher}{ACM}, \bibinfo{address}{New York}.
\newblock
\showISBNx{9798400703300}
\href{https://doi.org/10.1145/3613904.3642773}{doi:\nolinkurl{10.1145/3613904.3642773}}


\bibitem[Keuning et~al\mbox{.}(2019)]%
        {keuning_systematic_2019}
\bibfield{author}{\bibinfo{person}{Hieke Keuning}, \bibinfo{person}{Johan Jeuring}, {and} \bibinfo{person}{Bastiaan Heeren}.} \bibinfo{year}{2019}\natexlab{}.
\newblock \showarticletitle{A {Systematic} {Literature} {Review} of {Automated} {Feedback} {Generation} for {Programming} {Exercises}}.
\newblock \bibinfo{journal}{\emph{ACM Transactions on Computing Education}}  \bibinfo{volume}{19} (\bibinfo{year}{2019}), \bibinfo{pages}{1--43}.
\newblock
\showISSN{1946-6226}
\href{https://doi.org/10.1145/3231711}{doi:\nolinkurl{10.1145/3231711}}


\bibitem[Kiesler et~al\mbox{.}(2024)]%
        {kiesler2023exploring}
\bibfield{author}{\bibinfo{person}{Natalie Kiesler}, \bibinfo{person}{Dominic Lohr}, {and} \bibinfo{person}{Hieke Keuning}.} \bibinfo{year}{2024}\natexlab{}.
\newblock \showarticletitle{Exploring the Potential of Large Language Models to Generate Formative Programming Feedback}. In \bibinfo{booktitle}{\emph{2023 IEEE Frontiers in Education Conference}}.
\newblock
\href{https://doi.org/10.1109/FIE58773.2023.10343457}{doi:\nolinkurl{10.1109/FIE58773.2023.10343457}}


\bibitem[Kiesler et~al\mbox{.}(2025)]%
        {kiesler2025therole}
\bibfield{author}{\bibinfo{person}{Natalie Kiesler}, \bibinfo{person}{Jacqueline Smith}, \bibinfo{person}{Juho Leinonen}, \bibinfo{person}{Armando Fox}, \bibinfo{person}{Stephen MacNeil}, {and} \bibinfo{person}{Petri Ihantola}.} \bibinfo{year}{2025}\natexlab{}.
\newblock \showarticletitle{The Role of Generative AI in Software Student CollaborAItion}. In \bibinfo{booktitle}{\emph{Proceedings of the 30th ACM Conference on Innovation and Technology in Computer Science Education V. 1}} \emph{(\bibinfo{series}{ITiCSE 2025})}. \bibinfo{publisher}{ACM}, \bibinfo{address}{New York}, \bibinfo{pages}{72–78}.
\newblock
\showISBNx{9798400715679}
\href{https://doi.org/10.1145/3724363.3729040}{doi:\nolinkurl{10.1145/3724363.3729040}}


\bibitem[Kuckartz and Rädiker(2019)]%
        {kuckartz_analyzing_2019}
\bibfield{author}{\bibinfo{person}{Udo Kuckartz} {and} \bibinfo{person}{Stefan Rädiker}.} \bibinfo{year}{2019}\natexlab{}.
\newblock \bibinfo{booktitle}{\emph{Analyzing {Qualitative} {Data} with {MAXQDA}: {Text}, {Audio}, and {Video}}}.
\newblock \bibinfo{publisher}{Springer International Publishing}, \bibinfo{address}{Cham}.
\newblock
\showISBNx{978-3-030-15670-1 978-3-030-15671-8}


\bibitem[Liffiton et~al\mbox{.}(2023)]%
        {liffiton_codehelp_2023}
\bibfield{author}{\bibinfo{person}{Mark Liffiton}, \bibinfo{person}{Brad Sheese}, \bibinfo{person}{Jaromir Savelka}, {and} \bibinfo{person}{Paul Denny}.} \bibinfo{year}{2023}\natexlab{}.
\newblock \showarticletitle{{CodeHelp}: {Using} {Large} {Language} {Models} with {Guardrails} for {Scalable} {Support} in {Programming} {Classes}}. In \bibinfo{booktitle}{\emph{Proceedings of the 23rd {Koli} {Calling} {International} {Conference} on {Computing} {Education} {Research}}}. \bibinfo{publisher}{ACM}.
\newblock
\href{https://doi.org/10.1145/3631802.3631830}{doi:\nolinkurl{10.1145/3631802.3631830}}


\bibitem[Lohr et~al\mbox{.}(2025)]%
        {lohr2025you}
\bibfield{author}{\bibinfo{person}{Dominic Lohr}, \bibinfo{person}{Hieke Keuning}, {and} \bibinfo{person}{Natalie Kiesler}.} \bibinfo{year}{2025}\natexlab{}.
\newblock \showarticletitle{You're (Not) My Type- Can LLMs Generate Feedback of Specific Types for Introductory Programming Tasks?}
\newblock \bibinfo{journal}{\emph{JCAL}} \bibinfo{volume}{41}, \bibinfo{number}{1} (\bibinfo{year}{2025}).
\newblock
\href{https://doi.org/10.1111/jcal.13107}{doi:\nolinkurl{10.1111/jcal.13107}}


\bibitem[Luxton-Reilly et~al\mbox{.}(2018)]%
        {luxtonreilly2018introductory}
\bibfield{author}{\bibinfo{person}{Andrew Luxton-Reilly}, \bibinfo{person}{Simon}, \bibinfo{person}{Ibrahim Albluwi}, \bibinfo{person}{Brett~A. Becker}, \bibinfo{person}{Michail Giannakos}, \bibinfo{person}{Amruth~N. Kumar}, \bibinfo{person}{Linda Ott}, \bibinfo{person}{James Paterson}, \bibinfo{person}{Michael~James Scott}, \bibinfo{person}{Judy Sheard}, {and} \bibinfo{person}{Claudia Szabo}.} \bibinfo{year}{2018}\natexlab{}.
\newblock \showarticletitle{Introductory programming: a systematic literature review}. In \bibinfo{booktitle}{\emph{ITiCSE 2018 Companion}}. \bibinfo{publisher}{ACM}, \bibinfo{address}{New York}, \bibinfo{pages}{55–106}.
\newblock
\showISBNx{9781450362238}
\href{https://doi.org/10.1145/3293881.3295779}{doi:\nolinkurl{10.1145/3293881.3295779}}


\bibitem[MacNeil et~al\mbox{.}(2023)]%
        {macneil2022experiences}
\bibfield{author}{\bibinfo{person}{Stephen MacNeil}, \bibinfo{person}{Andrew Tran}, \bibinfo{person}{Arto Hellas}, \bibinfo{person}{Joanne Kim}, \bibinfo{person}{Sami Sarsa}, \bibinfo{person}{Paul Denny}, \bibinfo{person}{Seth Bernstein}, {and} \bibinfo{person}{Juho Leinonen}.} \bibinfo{year}{2023}\natexlab{}.
\newblock \showarticletitle{{Experiences from Using Code Explanations Generated by Large Language Models in a Web Software Development E-Book}}. In \bibinfo{booktitle}{\emph{Proc. SIGCSE TS}}. \bibinfo{pages}{931–937}.
\newblock
\href{https://doi.org/10.1145/3545945.3569785}{doi:\nolinkurl{10.1145/3545945.3569785}}


\bibitem[Mahmood et~al\mbox{.}(2025)]%
        {mahmood_user_2025}
\bibfield{author}{\bibinfo{person}{Amama Mahmood}, \bibinfo{person}{Junxiang Wang}, \bibinfo{person}{Bingsheng Yao}, \bibinfo{person}{Dakuo Wang}, {and} \bibinfo{person}{Chien-Ming Huang}.} \bibinfo{year}{2025}\natexlab{}.
\newblock \showarticletitle{User {Interaction} {Patterns} and {Breakdowns} in {Conversing} with {LLM}-{Powered} {Voice} {Assistants}}.
\newblock \bibinfo{journal}{\emph{International Journal of Human-Computer Studies}}  \bibinfo{volume}{195} (\bibinfo{year}{2025}).
\newblock
\showISSN{1071-5819}
\href{https://doi.org/10.1016/j.ijhcs.2024.103406}{doi:\nolinkurl{10.1016/j.ijhcs.2024.103406}}


\bibitem[Narciss(2008)]%
        {narciss2008feedback}
\bibfield{author}{\bibinfo{person}{Susanne Narciss}.} \bibinfo{year}{2008}\natexlab{}.
\newblock \showarticletitle{Feedback strategies for interactive learning tasks}.
\newblock \bibinfo{journal}{\emph{Handbook of research on educational communications and technology}}  \bibinfo{volume}{3} (\bibinfo{year}{2008}), \bibinfo{pages}{125--144}.
\newblock


\bibitem[OpenAI(2024)]%
        {openai_gpt-4o_2024}
\bibfield{author}{\bibinfo{person}{OpenAI}.} \bibinfo{year}{2024}\natexlab{}.
\newblock \bibinfo{title}{{GPT}-4o {System} {Card}}.
\newblock
\href{https://doi.org/10.48550/arXiv.2410.21276}{doi:\nolinkurl{10.48550/arXiv.2410.21276}}


\bibitem[Phung et~al\mbox{.}(2023)]%
        {phung2023generating}
\bibfield{author}{\bibinfo{person}{Tung Phung}, \bibinfo{person}{José Cambronero}, \bibinfo{person}{Sumit Gulwani}, \bibinfo{person}{Tobias Kohn}, \bibinfo{person}{Rupak Majumdar}, \bibinfo{person}{Adish Singla}, {and} \bibinfo{person}{Gustavo Soares}.} \bibinfo{year}{2023}\natexlab{}.
\newblock \showarticletitle{Generating High-Precision Feedback for Programming Syntax Errors using Large Language Models}. In \bibinfo{booktitle}{\emph{Proceedings of the 16th International Conference on Educational Data Mining}}. \bibinfo{publisher}{International Educational Data Mining Society}, \bibinfo{pages}{370--377}.
\newblock
\href{https://doi.org/10.5281/zenodo.8115653}{doi:\nolinkurl{10.5281/zenodo.8115653}}


\bibitem[Poddar et~al\mbox{.}(2024)]%
        {poddar_experiences_2024}
\bibfield{author}{\bibinfo{person}{Roshni Poddar}, \bibinfo{person}{Tarini Naik}, \bibinfo{person}{Manikanteswar Punnam}, \bibinfo{person}{Kavyansh Chourasia}, \bibinfo{person}{Rajeswari Pandurangan}, \bibinfo{person}{Rajesh~S Paali}, \bibinfo{person}{Nagarathna~R Bhat}, \bibinfo{person}{Bhagyashree Biradar}, \bibinfo{person}{Venkatesh Deshpande}, \bibinfo{person}{Devidatta Ghosh}, \bibinfo{person}{Sudipta~Ray Chaudhuri}, \bibinfo{person}{Dipanjan Chakraborty}, \bibinfo{person}{Amit Prakash}, {and} \bibinfo{person}{Manohar Swaminathan}.} \bibinfo{year}{2024}\natexlab{}.
\newblock \showarticletitle{Experiences from {Running} a {Voice}-{Based} {Education} {Platform} for {Children} and {Teachers} with {Visual} {Impairments}}.
\newblock \bibinfo{journal}{\emph{ACM J. Comput. Sustain. Soc.}} \bibinfo{volume}{2}, \bibinfo{number}{3} (\bibinfo{year}{2024}), \bibinfo{pages}{1--35}.
\newblock
\showISSN{2834-5533}
\href{https://doi.org/10.1145/3677323}{doi:\nolinkurl{10.1145/3677323}}


\bibitem[Prather et~al\mbox{.}(2023a)]%
        {prather2023therobots:wgfull}
\bibfield{author}{\bibinfo{person}{James Prather}, \bibinfo{person}{Paul Denny}, \bibinfo{person}{Juho Leinonen}, \bibinfo{person}{Brett~A. Becker}, \bibinfo{person}{Ibrahim Albluwi}, \bibinfo{person}{Michelle Craig}, \bibinfo{person}{Hieke Keuning}, \bibinfo{person}{Natalie Kiesler}, \bibinfo{person}{Tobias Kohn}, \bibinfo{person}{Andrew Luxton-Reilly}, \bibinfo{person}{Stephen MacNeil}, \bibinfo{person}{Andrew Petersen}, \bibinfo{person}{Raymond Pettit}, \bibinfo{person}{Brent~N. Reeves}, {and} \bibinfo{person}{Jaromir Savelka}.} \bibinfo{year}{2023}\natexlab{a}.
\newblock \showarticletitle{The Robots Are Here: Navigating the Generative AI Revolution in Computing Education}. In \bibinfo{booktitle}{\emph{Proceedings of the 2023 Working Group Reports on Innovation and Technology in Computer Science Education}}. \bibinfo{publisher}{ACM}, \bibinfo{address}{New York}, \bibinfo{pages}{108–159}.
\newblock
\showISBNx{9798400704055}
\href{https://doi.org/10.1145/3623762.3633499}{doi:\nolinkurl{10.1145/3623762.3633499}}


\bibitem[Prather et~al\mbox{.}(2025)]%
        {prather_beyond_2024}
\bibfield{author}{\bibinfo{person}{James Prather}, \bibinfo{person}{Juho Leinonen}, \bibinfo{person}{Natalie Kiesler}, \bibinfo{person}{Jamie Gorson~Benario}, \bibinfo{person}{Sam Lau}, \bibinfo{person}{Stephen MacNeil}, \bibinfo{person}{Narges Norouzi}, \bibinfo{person}{Simone Opel}, \bibinfo{person}{Vee Pettit}, \bibinfo{person}{Leo Porter}, \bibinfo{person}{Brent~N. Reeves}, \bibinfo{person}{Jaromir Savelka}, \bibinfo{person}{David~H. Smith}, \bibinfo{person}{Sven Strickroth}, {and} \bibinfo{person}{Daniel Zingaro}.} \bibinfo{year}{2025}\natexlab{}.
\newblock \showarticletitle{Beyond the Hype: A Comprehensive Review of Current Trends in Generative AI Research, Teaching Practices, and Tools}. In \bibinfo{booktitle}{\emph{2024 Working Group Reports on Innovation and Technology in Computer Science Education}}. \bibinfo{publisher}{ACM}, \bibinfo{address}{New York}, \bibinfo{pages}{300–338}.
\newblock
\showISBNx{9798400712081}
\href{https://doi.org/10.1145/3689187.3709614}{doi:\nolinkurl{10.1145/3689187.3709614}}


\bibitem[Prather et~al\mbox{.}(2018)]%
        {prather_metacognitive_2018}
\bibfield{author}{\bibinfo{person}{James Prather}, \bibinfo{person}{Raymond Pettit}, \bibinfo{person}{Kayla McMurry}, \bibinfo{person}{Alani Peters}, \bibinfo{person}{John Homer}, {and} \bibinfo{person}{Maxine Cohen}.} \bibinfo{year}{2018}\natexlab{}.
\newblock \showarticletitle{Metacognitive {Difficulties} {Faced} by {Novice} {Programmers} in {Automated} {Assessment} {Tools}}. In \bibinfo{booktitle}{\emph{Proc. of the 2018 ICER}}. \bibinfo{publisher}{ACM}, \bibinfo{address}{New York}, \bibinfo{pages}{41--50}.
\newblock
\href{https://doi.org/10.1145/3230977.3230981}{doi:\nolinkurl{10.1145/3230977.3230981}}


\bibitem[Prather et~al\mbox{.}(2023b)]%
        {prather_its_2023}
\bibfield{author}{\bibinfo{person}{James Prather}, \bibinfo{person}{Brent~N. Reeves}, \bibinfo{person}{Paul Denny}, \bibinfo{person}{Brett~A. Becker}, \bibinfo{person}{Juho Leinonen}, \bibinfo{person}{Andrew Luxton-Reilly}, \bibinfo{person}{Garrett Powell}, \bibinfo{person}{James Finnie-Ansley}, {and} \bibinfo{person}{Eddie~Antonio Santos}.} \bibinfo{year}{2023}\natexlab{b}.
\newblock \showarticletitle{“It’s Weird That it Knows What I Want”: Usability and Interactions with Copilot for Novice Programmers}.
\newblock \bibinfo{journal}{\emph{ACM Trans. Comput.-Hum. Interact.}} \bibinfo{volume}{31}, \bibinfo{number}{1} (\bibinfo{year}{2023}), \bibinfo{numpages}{31}~pages.
\newblock
\showISSN{1073-0516}
\href{https://doi.org/10.1145/3617367}{doi:\nolinkurl{10.1145/3617367}}


\bibitem[Prather et~al\mbox{.}(2024)]%
        {prather_widening_2024}
\bibfield{author}{\bibinfo{person}{James Prather}, \bibinfo{person}{Brent~N Reeves}, \bibinfo{person}{Juho Leinonen}, \bibinfo{person}{Stephen MacNeil}, \bibinfo{person}{Arisoa~S Randrianasolo}, \bibinfo{person}{Brett~A. Becker}, \bibinfo{person}{Bailey Kimmel}, \bibinfo{person}{Jared Wright}, {and} \bibinfo{person}{Ben Briggs}.} \bibinfo{year}{2024}\natexlab{}.
\newblock \showarticletitle{The {Widening} {Gap}: {The} {Benefits} and {Harms} of {Generative} {AI} for {Novice} {Programmers}}. In \bibinfo{booktitle}{\emph{Proc. of 2024 ICER}}, Vol.~\bibinfo{volume}{1}. \bibinfo{publisher}{ACM}, \bibinfo{address}{New York}, \bibinfo{pages}{469--486}.
\newblock
\showISBNx{979-8-4007-0475-8}
\href{https://doi.org/10.1145/3632620.3671116}{doi:\nolinkurl{10.1145/3632620.3671116}}


\bibitem[Radford et~al\mbox{.}(2022)]%
        {radford_robust_2022}
\bibfield{author}{\bibinfo{person}{Alec Radford}, \bibinfo{person}{Jong~Wook Kim}, \bibinfo{person}{Tao Xu}, \bibinfo{person}{Greg Brockman}, \bibinfo{person}{Christine McLeavey}, {and} \bibinfo{person}{Ilya Sutskever}.} \bibinfo{year}{2022}\natexlab{}.
\newblock \bibinfo{title}{Robust {Speech} {Recognition} via {Large}-{Scale} {Weak} {Supervision}}.
\newblock
\href{https://doi.org/10.48550/arXiv.2212.04356}{doi:\nolinkurl{10.48550/arXiv.2212.04356}}


\bibitem[Sayago(2021)]%
        {sayago_voice_2021}
\bibfield{author}{\bibinfo{person}{Sergio Sayago}.} \bibinfo{year}{2021}\natexlab{}.
\newblock \showarticletitle{Voice {Assistants} as {Learning} {Companions}}. In \bibinfo{booktitle}{\emph{22nd {International} {Conference} on {Human}-{Computer} {Interaction} with {Mobile} {Devices} and {Services}}}. \bibinfo{publisher}{ACM}, \bibinfo{address}{New York}, \bibinfo{pages}{1--3}.
\newblock
\showISBNx{978-1-4503-8052-2}
\href{https://doi.org/10.1145/3406324.3410707}{doi:\nolinkurl{10.1145/3406324.3410707}}


\bibitem[Scholl and Kiesler(2024)]%
        {scholl_how_2024}
\bibfield{author}{\bibinfo{person}{Andreas Scholl} {and} \bibinfo{person}{Natalie Kiesler}.} \bibinfo{year}{2024}\natexlab{}.
\newblock \showarticletitle{How {Novice} {Programmers} {Use} and {Experience} {ChatGPT} when {Solving} {Programming} {Exercises} in an {Introductory} {Course}}. In \bibinfo{booktitle}{\emph{2024 {IEEE} {Frontiers} in {Education} {Conference} ({FIE})}}. \bibinfo{publisher}{IEEE}, \bibinfo{address}{Washington, DC, USA}, \bibinfo{pages}{1--9}.
\newblock
\showISBNx{979-8-3503-5150-7}
\href{https://doi.org/10.1109/FIE61694.2024.10893442}{doi:\nolinkurl{10.1109/FIE61694.2024.10893442}}


\bibitem[Scholl and Kiesler(2025)]%
        {scholl2025students}
\bibfield{author}{\bibinfo{person}{Andreas Scholl} {and} \bibinfo{person}{Natalie Kiesler}.} \bibinfo{year}{2025}\natexlab{}.
\newblock \bibinfo{title}{Students' Feedback Requests and Interactions with the SCRIPT Chatbot: Do They Get What They Ask For?}
\newblock
\href{https://doi.org/10.48550/arXiv.2507.17258}{doi:\nolinkurl{10.48550/arXiv.2507.17258}}
\newblock
\shownote{PPIG 2025 Workshop}.


\bibitem[Scholl et~al\mbox{.}(2024)]%
        {scholl_analyzing_2024}
\bibfield{author}{\bibinfo{person}{Andreas Scholl}, \bibinfo{person}{Daniel Schiffner}, {and} \bibinfo{person}{Natalie Kiesler}.} \bibinfo{year}{2024}\natexlab{}.
\newblock \showarticletitle{Analyzing {{Chat Protocols}} of {{Novice Programmers Solving Introductory Programming Tasks}} with {{ChatGPT}}}. In \bibinfo{booktitle}{\emph{Proceedings of DELFI 2024}}, \bibfield{editor}{\bibinfo{person}{Sandra Schulz} {and} \bibinfo{person}{Natalie Kiesler}} (Eds.). \bibinfo{publisher}{Gesellschaft f{\"u}r Informatik e.V.}
\newblock
\href{https://doi.org/10.18420/delfi2024_05}{doi:\nolinkurl{10.18420/delfi2024_05}}


\bibitem[Seaborn et~al\mbox{.}(2024)]%
        {seaborn_cross-cultural_2024}
\bibfield{author}{\bibinfo{person}{Katie Seaborn}, \bibinfo{person}{Iona Gessinger}, \bibinfo{person}{Suzuka Yoshida}, \bibinfo{person}{Benjamin~R. Cowan}, {and} \bibinfo{person}{Philip~R. Doyle}.} \bibinfo{year}{2024}\natexlab{}.
\newblock \showarticletitle{Cross-{Cultural} {Validation} of {Partner} {Models} for {Voice} {User} {Interfaces}}. In \bibinfo{booktitle}{\emph{Proceedings of the 6th {ACM} {Conference} on {Conversational} {User} {Interfaces}}} \emph{(\bibinfo{series}{{CUI} '24})}. \bibinfo{publisher}{ACM}, \bibinfo{address}{New York}, \bibinfo{pages}{1--10}.
\newblock
\showISBNx{979-8-4007-0511-3}
\href{https://doi.org/10.1145/3640794.3665537}{doi:\nolinkurl{10.1145/3640794.3665537}}


\bibitem[Seaborn et~al\mbox{.}(2025)]%
        {Seaborn2025data}
\bibfield{author}{\bibinfo{person}{Katie Seaborn}, \bibinfo{person}{Suzuka Yoshida}, \bibinfo{person}{Philip Doyle}, \bibinfo{person}{Iona Gessinger}, \bibinfo{person}{Benjamin~R Cowan}, {and} \bibinfo{person}{Nazlinur Gokturk}.} \bibinfo{year}{2025}\natexlab{}.
\newblock \bibinfo{title}{Localizing the PMQ}.
\newblock
\href{https://doi.org/10.17605/OSF.IO/GN6WJ}{doi:\nolinkurl{10.17605/OSF.IO/GN6WJ}}


\bibitem[Stone(2024)]%
        {stone_exploring_2024}
\bibfield{author}{\bibinfo{person}{Irene Stone}.} \bibinfo{year}{2024}\natexlab{}.
\newblock \showarticletitle{Exploring {Human}-{Centered} {Approaches} in {Generative} {AI} and {Introductory} {Programming} {Research}: {A} {Scoping} {Review}}. In \bibinfo{booktitle}{\emph{Proceedings of the 2024 {Conference} on {United} {Kingdom} \& {Ireland} {Computing} {Education} {Research}}}. \bibinfo{publisher}{ACM}, \bibinfo{pages}{1--7}.
\newblock
\showISBNx{979-8-4007-1177-0}
\href{https://doi.org/10.1145/3689535.3689553}{doi:\nolinkurl{10.1145/3689535.3689553}}


\bibitem[Yin(2017)]%
        {yin_case_2017}
\bibfield{author}{\bibinfo{person}{Robert~K. Yin}.} \bibinfo{year}{2017}\natexlab{}.
\newblock \bibinfo{booktitle}{\emph{Case {Study} {Research} and {Applications}: {Design} and {Methods}}}.
\newblock \bibinfo{publisher}{SAGE Publications}.
\newblock
\showISBNx{978-1-5063-3618-3}


\end{thebibliography}

\end{document}